# CloudPass – a passport system based on Cloud Computing and Near Field Communication

Adethya Sudarsanan, Programmer Analyst, Cognizant Technology Solutions India Pvt. Ltd.

E-mail: adethyasjce@gmail.com

**Abstract**

Wireless communication has penetrated into all fields of technology, especially in mobility, where wireless transactions are gaining importance with improvements in standards like 3G and 4G. There are many technologies that support the wireless forms of interactions between devices. One among them is NFC – Near Field Communication. In addition to NFC, other external technologies like Quick Response (QR) Codes assist in establishing interactions among participating devices. In this paper, we examine an approach that will involve standards and technologies like NFC, QR Codes and Cloud Infrastructure to design a mobile application which will perform desired functionalities. Cloud Storage is used as a reservoir to store the artifacts used by the application. Development and testing of the application is initially carried out on emulators or simulators followed by testing on real handsets/devices.

**Near Field Communication (NFC)**

Near Field Communication (NFC) is a set of standards for smart phones and similar devices to establish radio communication with each other by touching them together or bringing them into close proximity, usually no more than 15 centimeters [1].

NFC operates on communication protocols and data exchange formats which are based on Radio Frequency Identification (RFID) standards including ISO/IEC 14443 and FeliCa and ISO/IEC 18092. NFC is basically built upon the existing RFID concepts by allowing two way communication exchanges between participating entities. NFC devices are also capable of replacing contact-less smart cards, which are examples of one way communication.

**Quick Response Code (QR Codes)**

Quick Response Codes or QR Codes are primarily used for storing information in encrypted form. These codes are designed to facilitate quick and faster decoding of the information stored in them. These codes support high speed decryption. The codes are represented as black modules arranged in a square pattern on a white background.

The information encrypted in a QR code can be made up of four standardized modes [2] of data such as numeric, alphanumeric, byte or binary and Kanji. In addition to these four modes, combinations of any two modes are also possible for encryption.

**Cloud Computing and Storage**

Cloud Service is the delivery of services over the internet. The services delivered involve a wide range of traditional services or products like desktops, software, security features, storage etc. and also new services to match the growing needs of the industry and people.

Cloud Storage is one of the cloud services which enables clients to access a storage space over the internet. This system involves providing designated storage space to clients on a cloud. This space can be accessed by the client to store information and other data which can be accessed by the client only.

**Existing System**

Passport is an official travel document issued by a country's government to its citizens. Along with a valid passport, the user should also possess a document to enter the destination country. This document is called Visa. Visas are usually issued as stickers which are affixed to the passport. These essential travel documents need to be carried every time one travels to another country.

The following are the limitations of the existing system

1. Obtaining a visa is a time consuming process
2. There is a need to deposit the passport with the embassy for visa stamping.

3. During immigration checks, the user's passport and visa is subjected to manual checking which consumes lot of time
4. There is every possibility of passport being duplicated or counterfeited.
5. Care has to be exercised to ensure that the user holds it safely and does not lose it.
6. Passport along with visa should be carried every time one travels.
7. Passport, as it is made of paper is liable to normal wear and tear, damage by dust, water, fire etc.
8. Passport will expire after stipulated number of years, and needs to be renewed to use it again.
9. More time is taken when there is any discrepancy in the user's passport or visa.

**Proposed System**

The proposed system suggests that traditional passport is replaced by an application that will act as passport. This application is designed to work on all devices and all platforms.

*Current target devices*: iPhone, iPad, iPod

*Current target platform*: Apple iOS

The following is the proposed idea

1. The user does not have a traditional passport. He simple owns an app on his apple device (let's say, iPhone 4).
2. The user downloads the passport app from appstore and installs it in his handheld device
3. The app acts as a passport when the user travels abroad.
4. Visa can be obtained as an image and the same can be placed inside the passport app.
5. Once the visa image is placed in the passport app, the user is ready to use his handheld device as a passport with a valid travel visa in it.
6. At the immigration desk, the user flashes his device at a designated NFC reader counter and the visa and passport both are checked for authenticity.
7. If the user passes the NFC counter check, he is free to travel. In case he fails, his device will be locked and airport police will be alerted.
8. Upon landing at the destination country, the user again flashes his handheld at the NFC counter where date and time of arrival is stamped to his visa inside the passport app.
9. The visa image obtained at the immigration desk is validated against the visa image present in the airport cloud. This airport cloud obtains data from the embassy cloud on a daily basis.

**Architecture of proposed system**

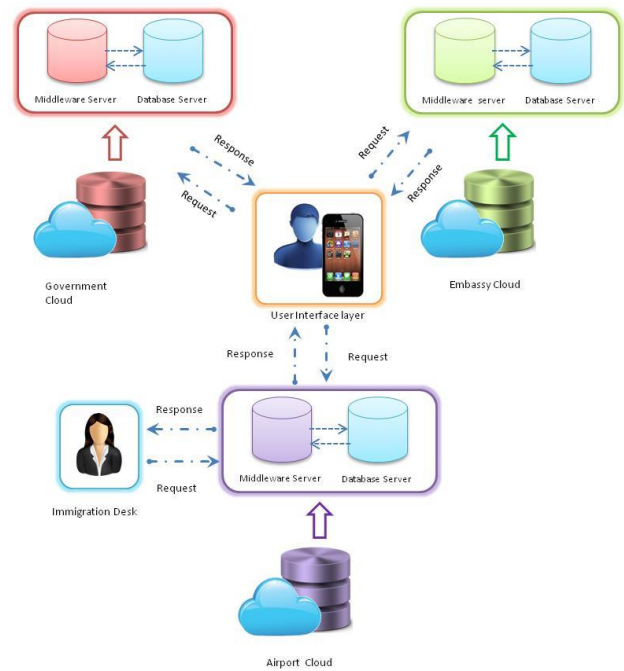

**Process Flow**

1. *User applies for and obtains passport*

    I. User logs into government website to apply for a passport online. He downloads the necessary forms and duly fills and submits them.
    II. A unique ID number is issued to the user upon submission, which will help him track the progress of his passport application.

  III. Once the passport is ready, the user is notified through e-mail. The e-mail contains a link from where the passport app can be downloaded.
  IV. This link directs to the cloud of passport issuing authority, which stores his passport as an application which can be downloaded into his handheld device.

2. *User applies for visa*

  I. User logs into destination country's visa issuing website to apply for a visa online. He downloads the necessary forms and duly fills and submits them.
  II. A unique ID number is issued to the user upon submission, which will help him track the progress of his visa application.
  III. Once the visa is ready, the user is notified through e-mail. The e-mail contains a QR Code from where the visa image can be downloaded.
  IV. This QR Code directs to the cloud of visa issuing authority, which stores his visa as an image which can be downloaded into the user's passport application.
  V. In addition to the downloading facility, the user also has the flexibility to choose the page in passport where the visa can be placed.

3. *User shows the passport app and visa image at the immigration desk (departure & arrival country)*

  I. The visa image shown by the user is read and a copy of the visa image is sent to the Airport Cloud present in the airport premises.
  II. The airport cloud is updated with visa images of all those travelling through that airport. This update is done from the respective Embassy Cloud.
  III. The visa obtained from the user and the visa obtained from the Embassy Cloud is compared to check if they are the same.
  IV. If they are same, the user is either permitted to board the aircraft from his departure location or allowed entry into the country of arrival. In case there is any discrepancy, the user is isolated for further security checks.

**Security mechanisms in the proposed system**

Security is always a matter of concern when traditional systems are moved into IT infrastructure. In this case we provide two levels of authentication.

The primary step in authentication is to make sure that the user is not an automated system rather a human who is trying to login into the application.

**Level 1 – time based authentication**

1. Time based authentication revolves around the concept where the user is expected to enter the current time displayed in his handheld along with a visual captcha.
2. Time does not reflect the time zone variations.

E.g. let's assume that a user travels to America with a device which is currently showing Indian Standard Time. Even after going to USA, the device will show only Indian time. So the displayed Indian time will be used for authentication. If the user is asked to enter time based authentication, he should enter the time displayed in his device without reflecting time zone differences.

*Steps to perform Time Based Authentication*

1. The user opens the passport app on his device.
2. The application requires the user to enter his current time and visual captcha displayed on his handheld.
3. The time and captcha is sent to the server for authentication.

4. Once the authentication is verified, the application prompts the user to enter his username and password.
5. This is again sent to the server for authentication.
6. Once the authentication is passed, only the passport of the user is displayed.
7. To display his visa, the user is required to go through another level of authentication called "Image Based Authentication"

The first time based check is made to ensure that human is accessing the application. The second check is made to ensure that the user is correct one.

**Level 2 – Image based authentication**

1. Image based authentication works on the principle that the user has to enter an answer for the image shown on the screen.
2. There are 10 images stored in the application, which are shown to the user in random fashion.
3. The user is expected to remember answers for all the 10 images. These images DO NOT change. Hence it is relatively easy for the user to remember 10 answers.
4. If the user enters the correct answer related to the image shown, he is given access to his visa.
5. This visa is shown at the immigration desk for further processing.

**Application Timeout**

The above authentication mechanisms will be active for 10 minutes from the time of activation. After 10 minutes, the application logs out automatically. This is done to ensure that the application's security is not compromised after the security check process.

**OTP Approach**

OTP – One Time Password mechanism is also involved in the security of this application. The password entered to perform a check expires when one transaction is over [3].

If for some reason, the user failed his security check but wants to try again, he can do that by entering the password authentication process completely again. This feature is enabled to ensure that application's security is not compromised in the time between one transaction and time out duration.

**Advantages of the Proposed System**

1. The proposed system is more secure than the traditional passport because it involves more security features like time and image authentication.
2. There is no duplication involved in the proposed system because the passport is supplied as an application which can be downloaded only after scanning the Quick Response Code.
3. NFC's are fast mechanisms which will ensure that the visa and passport verification are done quickly.
4. Time taken for verification will be greatly reduced and also the verification will be of greater accuracy.
5. As the passport and visa are applications installed on a device, no paper based passports are used.
6. There is no need to submit the passport for Visa stamping.
7. The proposed system is cost effective. Traditional passports and visa consume more money when compared to the proposed approach.

**Limitations of the Proposed System**

1. The user is expected to carry at least one handheld device with required features to support the application
2. Handling devices like NFC Reader needs the assistance of a trained professional
3. As traditional passports and visa are famous among the public, adapting to the proposed idea will take some time
4. Battery life, network connectivity, proper functioning of sensors are a few things to be taken into consideration when implementing the proposed system